# Surface Passivation of GaAs using Chemical and Plasma Methods

# D. Alexiev[1], D. A. Prokopovich[1,2], L. Mo[1]

[1]Australian Nuclear Science and Technology Organisation (ANSTO), PMB 1 Menai NSW 2234 Australia
[2]University of Wollongong, Wollongong, N.S.W. 2522 Australia

## ABSTRACT
Passivation of the GaAs surface was attempted using aqueous $P_2S_5$-$NH_4OH$, $(NH_4)_2S_x$ and plasma nitrogenataion and hydrogenation.
Results indicate that plasma nitrogenation with pre-treatment of plasma hydrogenation produced consistent reduction in reverse leakage current at room-temperature for all p and n type schottky diodes. Some diodes showed an order of magnitude improvement in current density.
Aqueous passivation showed similarly an improvement however, additional experimentation is required into long term stability and the arsenic-sulphur covalent bond strength.

## 1.1 INTRODUCTION

It would be of interest to look at the transport mechanism of the p-n junction under reverse bias before examining experimental Schottky diodes with various surface passivants. The leakage current $I_R$ of a reverse biased junction can be expressed as a sum of six current components:

$$I_R = I_d + I_{gb} + I_{gs} + I_{AV} + I_p + I_t$$

Where   $I_d$    = diffusion current
        $I_{gb}$ = bulk generation - recombination current
        $I_{gs}$ = surface generation - recombination current
        $I_{AV}$ = surface exchange current
        $I_p$    = current through the passivant
        $I_t$    = tunnel current through the passivant

The current component I is proportional to the minority-carrier d density outside the depletion region and should remain constant for applied bias $V_R$ greater than a few kT. Its temperature dependence should show an activation energy $E_a$ that equals bandgap $E_g$ of the semiconductor material. The currents $I_{gb}$ and $I_{gs}$ are proportional to their respective depletion widths and, therefore, have a weak VR dependence of $VR^{-1/3}$ to $V_R^{-1/2}$ for a diffused junction. $I_{AV}$ depends on the region near the junction and will depend on the level of interface state densities. Currents $I_{gs}$ and $I_{AV}$ can be altered by surface passivation; ideally, the passivant must be such that it will produce flat band conditions. $I_p$ and $I_t$ are also affected by the surface passivation.

Current $I_p$ should be negligible for high resistance passivants. The tunnelling current ($I_t$) must depend on $V_R$. Tunnelling can also occur through a combination of thermal excitation and field emission through gap states. The activation energy will then depend on the energy distribution of the gap states; it will be higher for transitions that are mostly through deep states than those that are mostly through shallow states. Under reverse bias, electrons from generation-recombination centres can tunnel out to the conduction band of the semiconductor material, but cannot go to the conduction band of the passivant because of the higher energy required.

Experimental radiation detectors (diodes) described by Alexiev et al, 1992 showed in general poor $V_R$ versus $I_R$ characteristic (Fig3) as often is the case, such characteristics are determined by the surface state densities (due to $I_{gs}$), impurities such as salt left after the rinse in deionized water, or by surface damage caused by earlier shaping and polishing of the crystal; not least is the surface discontinuity itself giving rise to dangling bonds. There is an obvious appeal in finding a chemical solution to such a problem, in particular, atomic hydrogenation of the surface. However, surface states, although arising from the breaking of specific bonds are delocalised over several atomic layers implying that even though a passivant may complete a bond, optimisation of the passivant species may be required.

In GaAs the root cause of poor electronic surfaces (Offsey et al (1986)) is the high density of surface states formed by segregated arsenic atoms (Spicer et al (1979)) through an oxidation reaction. This results in the Fermi level at the GaAs (100) surface being "pinned" at 0.8 eV below the conduction band. About 10 states cm are involved (Offsey et al (1985)) and several nonstoichiometric phases are present, some of which are conducting (Wieder (1980)). These phases consist of $As_2O_3$ and $Ga_2O_3$, $As_2O_3$ will be unstable because it reacts with Ga (in the vicinity of the arsenic oxide) by gradually extracting oxygen to form $Ga_2O_3$, leading to As segregation (Thermond et al (1980), Wilmsen (1982)) via the solid state reaction $2GaAs + As_2O_2 \longrightarrow Ga_2O_3 + 4As$. To counteract this process, Lee et al (1989) suggest that a successful passivant must be a species that makes the surface repel approaching oxygen and at the same time have a higher heat of oxide formation than that of gallium oxide. A further requirement must be that the selected species adsorbs strongly on the GaAs (100) surface as an impenetrable barrier. Such species, proposed by Lee et al, are phosphorous compounds. Lee et al selected in part $P_2S_5$, it has a higher heat of formation than gallium oxide and forms an impenetrable barrier to approaching oxygen.

Other efficient surface passivation techniques rely on a reaction between sulphides and the GaAs surface. These reactions form $As_xS_y$ phases which grow only when the oxide is stripped from the GaAs surface. Sandroff et al (1989) noted that one phase (As-S) forms as bulk $As_2S_y$ whilst another shows an in-plane S-S bonding. Sandroff et al also note that band bending on such a disulphide terminated surface is reduced to 0.12 eV. Similar results have been reported by Carpenter et al (1989), Cowens et al (1989) and Lee et al (1989). Sandroff et al notes the surface improvement gained when using $Na_2S.9H_2O$ persists only for a few days before reverting to its original state. This result was supported by Carpenter et al who found that Schottky barriers formed on ammonium sulphide $(NH_4)_2S$ treated n and p type GaAs showed no apparent aging after several months exposure to room air. Surface state density was reduced by an order of magnitude when compared to untreated GaAs. Similar results were reported by Fan et al (1989), attaining an interface density of about $1.2 \times 10^{11}$ cm$^{-2}$.eV$^{-1}$ for $(NH_4)_2S_x$ treated surfaces.

GaAs surface preparation usually follows a similar approach: ultrasonic cleaning in an organic solvent followed by an alkaline etch, typically $NH_4OH : H_2O_2 : H_2O = 3:2:95$ at 40°C for about 30 sec. Samples are then immediately treated with $(NH_4)_2S_x$ or $Na_2S.9H_2O$.

Other approaches to GaAs surface passivation (Ives et al (1987, 1990) and Kirschner (1988)) involved exposing the surface to flowing water and light in a photowash cell. They succeeded in unpinning or pinning the Fermi level as the oxygen content of the water was altered; when the water was saturated with hydrogen, large periods of band flattening were noted. The introduction by Ives et al of molecular species of hydrogen into the photowash suggest that some ionic bonding to the arsenic atom may occur with the following reaction: $GaAs + 3H_2O \longrightarrow Ga(OH)_3 + AsH_3$, which implies the formation of Ga-OH and As-H bonds. Such As-H bonds have been reported by Landesman et al (1989) when the GaAs surface is exposed to atomic hydrogen produced by a hydrogen plasma.

Plasma hydrogenation of GaAs defects, impurities and a number of deep centres has been studied over the past ten years (Tavendale et al (1980), Chevallier et al (1985), Johnson et al (1986) and Pearton et al (1987)). Shallow donors and acceptors and a number of deep centres can be passivated by atomic hydrogen and generally their electrical activity can be restored at higher temperatures. However, only little attention was paid to passivating the GaAs surface or the metal-surface interface with atomic hydrogen. First to publish such investigations was Callegari et al (1988); they found, after atomic hydrogenation of metal-oxide-semiconductor (MOS) capacitors, improvements in high-frequency response and C-V characteristics. Interface state density was reduced to about $10^{11}$ cm$^{-2}$.ev$^{-1}$ with longer term stability. Paccagnetta et al (1989) applied, a similar process to Schottky diodes finding an improvement in the rectifying properties with possible passivation of donors and bulk defects. However, only a partial reduction of the surface state density was noted, implying a partial Fermi level unpinning. The reason for such a poor surface passivation result can be found in a report by Omeljanovsky et al (1989) which observes that RF (tens kHz to tens MHz) generated atomic hydrogen species has the disadvantage of non-resonant interaction in the RF electric field with charged particles, consequently a low concentration of atomic hydrogen is produced. Their approach was to use a microwave (MCW) plasma source with a resonant magnetic plasma holding capacity (Balmashnov et al (1987)). The ionised hydrogen density was found to be two orders of magnitude higher (~$10^{12}$ cm$^{-3}$) than a conventional RF plasma. Omeljanovsky et al (their figures 12 and 13) achieved highly impressive I-V characteristics for hydrogenated Au barrier n-type GaAs Schottky diodes and a two order reduction in surface state density (~$10^{10}$ cm$^{-2}$.ev$^{-1}$). This result suggests that the results of Paccagnella et al could have been better had the (RF) atomic hydrogen (H$^+$) species been at a higher density. On the other hand, a high RF induced density could also produce surface damage due to sputtering and a compounding species of hydrogen, H$^+$ and H° (neutral), whose ratio depends on RF frequency and power input (private communication, Tavendale 1989). Consequently, low levels of H$^+$ are produced for passivation.

## 2.1 Experimental - Hydrogen Plasma (RF) Passivation

Even though Omeljanovsky et al indicate that RF hydrogen plasma for surface passivation has some inherent disadvantages, a limited set of experiments were performed using RF plasma. To achieve this, a simple arrangement using an 800 watt 27.5 MHz RF generator, was developed as shown schematically in Figure 2-1. Briefly, molecular hydrogen is supplied from a palladium diffuser into a silica column containing a heater pedestal topped with a CZ GaAs wafer which is used to hold and heat the LPE GaAs test samples. The CZ wafer and heater pedestal is shrouded with a silica tube to avoid sputtering contamination back onto the samples. A partial pressure of H was maintained with a vacuum system consisting of a rotary pump, a 120 $l.s^{-1}$ turbo molecular pump and a throttle valve located just below the heater pedestal. The heater can be controlled from ambient to 400°C at a resolution of 0.5°C using a simple ON-OFF type controller, so chosen to be immune to RF interference.

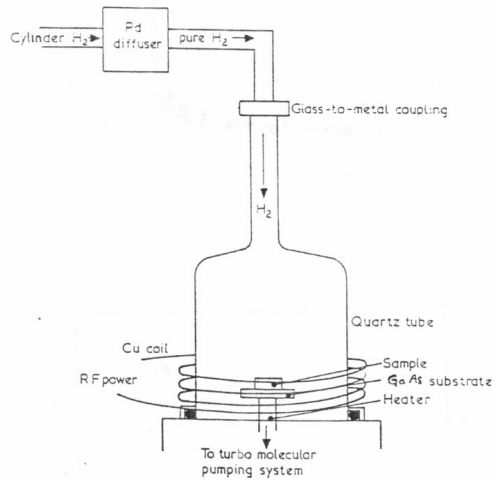

Figure 2-1. Schematic of Hydrogen Plasma system for incorporation of $H^+$ in semiconductors.
[Note: Cu coil can be moved up or down quartz tube, thereby altering the position of the plasma in respect to the GaAs samples].

All samples, about 5 mm square, were cleaned in an organic solvent followed by a 30 sec etch in $H_2SO_4 : H_2O_2 : H_2O$ = 3:1:1 AT 60 °C. After rinsing in deionised water, the samples were immediately loaded onto the heater pedestal for atomic hydrogenation. The hydrogen plasma was maintained about 6cm above the sample surfaces and at no time was it allowed to strike the surface. Sample temperatures used were 25°C, 175°C and 270°C. I-V and C-V measurements were done before and after hydrogen plasma exposure following standard metalisation.

Typically, before hydrogenation, such diodes displayed poor leakage currents. Leakage current activation energies were in the order of 0.75 to 0.8 eV, considerably less than the ideal value of 1.43 eV indicative of surface currents. Exposure to atomic hydrogen produced an improvement only when the samples were heated at 270°C and exposed for 2 hours to the plasma. However, some samples showed no change at all. Those that did improve with hydrogenation (typically as shown in Figure 2-2) were found to revert after a few days to their original leakage current, suggesting that the GaAs surface was only partially passivated and the Fermi level reverted to a pinned mid band position. These results confirm the comments of Omeljenovsky et al on the suitability of RF plasma hydrogenation for surface passivation.

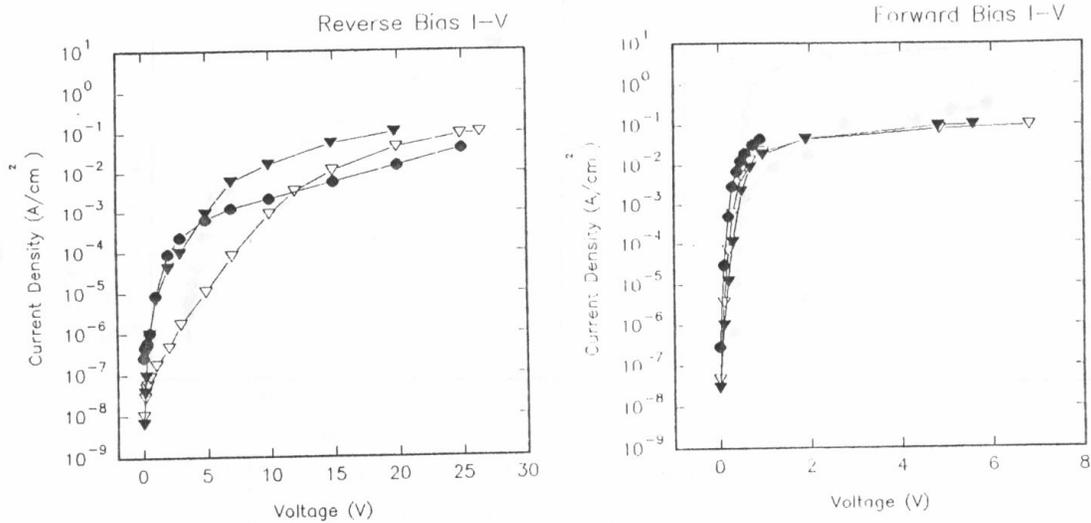

Figure 2-2. I-V characteristics of hydrogenated ($H^+$) LPE GaAs, showing margin improvement to the reverse current.

● etched in $H_2SO_4 : H_2O_2 : H_2O = 3:1:1$   ▽ hydrogenated ($H^+$), ▼ 3 days after $H^+$

## 2.2 <u>Nitrogen Plasma (R.F.) Passivation</u>

The usefulness of atomic hydrogen passivation of the (100) LPE-GaAs surface when the plasma is generated by RF has been found to produce only a marginal effect. However, when pretreating the GaAs surface with atomic hydrogen as described by Pearton et al (1984) and Callegari et al (1989) excess As which can be present either in elemental form or as an oxide $As_2O_3$ can be removed by the reaction:

As + 3H —> $AsH_3$ ↑
and     $As_2O_3$ + 12H —> $3H_2O$ ↑ + $2AsH_3$ ↑

Then, without further exposing the surface to oxidation, the excess Ga surface can be altered to form GaN a wide band gap (Eg = 3.5 eV) semiconductor layer.

The experimental approach was similar to 2.1. Briefly, each p and n type LPE GaAs section was etched in $H_2SO_4 : H_2O_2: H_2O = 3:1:1$ for 30 sec at 40°C, exposed to atomic hydrogen for approximately 0.5 H followed by 2 H exposure to atomic nitrogen. Sample temperatures were room temperature to 200°C for H+ and up to 360°C during $N^+$ treatment. After passivation all samples were metalised to form Schottky diodes and examined using standard characterisation techniques (I-V and C-V).

Results indicate that all p and n type Schottky diodes treated in this manner show a consistent reduction in leakage currents at room temperature with some diodes showing an order of magnitude improvement. The leakage current activation energy has also increased, from 0.6 eV to approximately 1 eV. Figure 2-3 shows the reverse current characteristic of an untreated sample and the same sample treated with N$^+$ RF plasma.

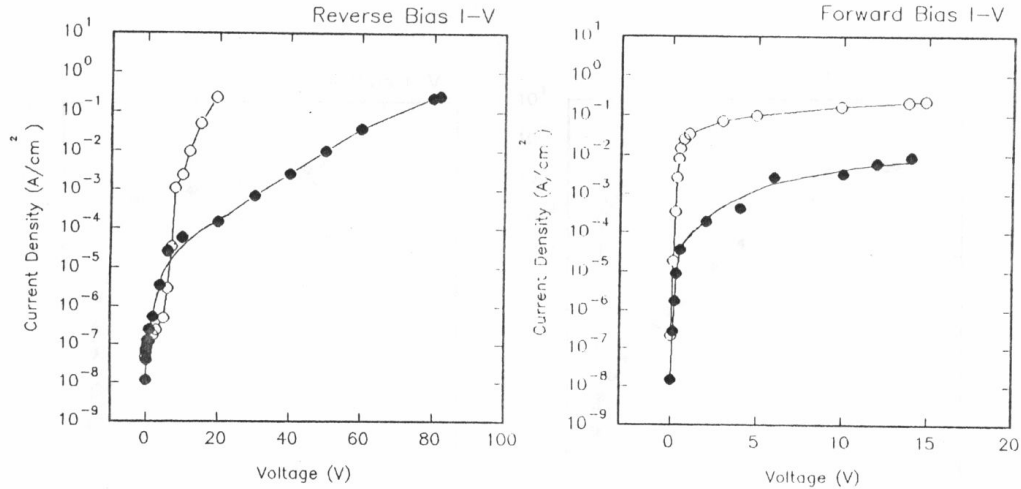

Figure 2-3. Reduction of leakage current due to nitrogen passivation of the (100) GaAs surface.
o etched in $H_2SO_4 : H_2O_2 : H_2O = 3:1:1$, ● nitridation (2H, 350°C).

## 3.1 The Aqueous $P_2S_5$ – $NH_4OH$ Surface Passivant

Lee et al (1989) claim that, when an n-type (100) GaAs surface is exposed to a solution of $PCl_3$ or $P_2S_5$ in $NH_4OH$, a fivefold increase in the photoluminescence can result, together with an improved current-voltage characteristic when the passivated sample is examined with an electrochemical Schottky contact. No other standard characterisation techniques were used. It is therefore of interest to repeat their preparation and examine Schottky diodes prepared with conventional techniques on such surfaces.

The experimental procedures were as follows: Samples used were both n and p type, low $10^{14}$ cm$^{-3}$ carrier density (100) LPE GaAs. The samples were cleaned in organic solvent and rinsed in methanol and deionised (18MΩ) water.

This was followed by a sequential series of alkaline based etchants:

$NH_4OH : H_2O_2 : H_2O = 3:2:95$;  4 min at about 40°C;
$NH_4OH : H_2O = 1:2$;  1 min at room temperature, used to remove a mixture of lower oxides left after the previous etch;

$HNO_3 : H_2O = 1:19$;  1 min at room temperature, used to remove elemental surface arsenic.

Finally, the samples were placed into the passivant solution:

$P_2S_5 : NH_4OH = 0.1$ g/ml at room temperature for various times.

After the final step, the samples were removed, dried with nitrogen and metallised in the normal way. All samples were then subjected to I-V and C-V profiling.

Results obtained show that all devices, p or n type, resulted in a poor (resistive) I-V characteristic (Figure 3-1). When plotting $I/C^2$ versus V (Figure 3-2), a clear reduction of barrier height for about 0.75 eV to near zero was noted. An attempt to affect the $P_2S_5$ treated surface by stripping any ionic bonded species such as $SO_3^-$, $SO_4^-$ with a reaction limited etchant ($NH_4OH : H_2O = 1:100$) did not alter the I-V and C-V characteristics. From the results observed it is clear that $P_2S_5/NH_4OH$ surface passivation of LPE GaAs is not useful for Schottky diode construction.

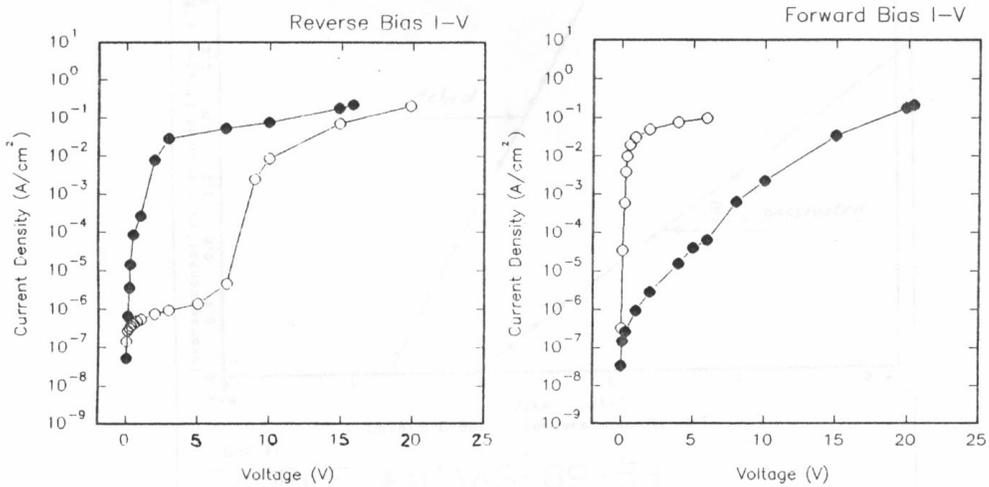

Figure 3-1. I-V characteristics for a (100) GaAs surface passivated with aqueous $P_2S_5$ - $NH_4OH$ solution.
○ alkaline etch, ● $P_2S_5$ passivation.

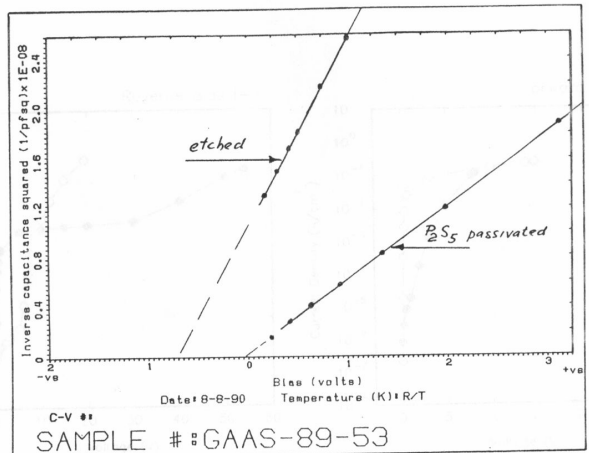

Figure 3-2. Reduction of barrier height from 0.75 eV to near zero eV when using the $P_2S_5$ - $NH_4OH$ passivant.

Perhaps it is clear that $P_2S_5$ passivation inserts a Chin resistive interlayer between GaAs and Schottky metallisation - so well passivated in fact that the metal fails to penetrate to the GaAs surface over most of the area. This would account for degraded final I-V characteristics, possibly higher reverse current due to local breakdown - but seems largely incompatible with the combination of larger capacitance and lower apparent barrier.

## 3.2 The Aqueous $(NH_4)_2S_x$ Passivant

The effects of ammonium sulphide surface treatment of (100) GaAs have been investigated mostly using X-ray photoelectron spectroscopy (XPS) and Raman scattering (Sandroff et al 1989). This involves comparing the magnitude of two related Raman spectra peaks labelled L⁻ and LO, corresponding to the degree of optical phonon scattering in the bulk and depleted regions, respectively. Sandroff et al found that in untreated samples, high density of surface states lead to a large depletion region and consequently the LO peak appears similar in intensity to the L" peak. Therefore, the degree of passivation, the reduction of surface state density, lead to a reduction of peak LO. To find the nature of peak LO, Sandroff et al used XPF, finding that the GaAs treated with $(NH_4)_2S_x$ has complete absence of the characteristic $As_2O_3$ peak; indicating that the LO peak is directly related to the oxide phase formation on the GaAs surface. The $(NH_4)_2S_x$ passivant has been shown to substantially reduce the intensity of peak LO. Carpenter et al (1989) investigated Schottky barrier formation on $(NH_4)_2S_x$ treated n and p-type GaAs, finding an increase in barrier height and a sensitivity to the metal work function. It is therefore of interest to verify this result by similarly treating LPE-GaAs with $(NH_4)_2S_x$, metallising and profiling for I-V and C-V characteristics.

The experimental procedure was similar to 3.1, after organic solvent cleaning all samples were etched, firstly in an alkaline based etchant with a second group of samples in a sulphuric based etchant. The GaAs samples were then placed for various time limits up to 5 hours, into the $(NH_4)_2S_x$ solution at 60°C. Again, after metallisation, I-V and C-V measurements were made. Figure 3-3 shows a very significant improvement in the reverse current characteristics. C-V measurements confirm an increase in barrier height.

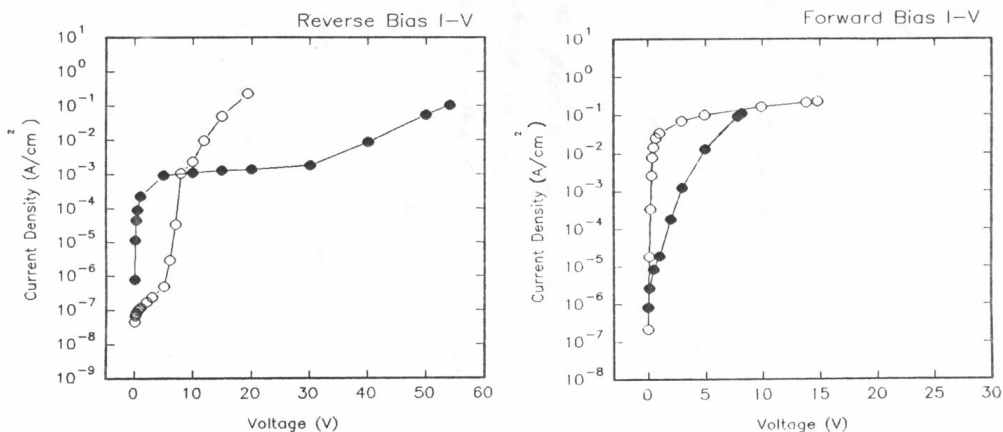

Figure 3-3. Shows a significant improvement in the reverse current characteristic when the (100) GaAs surface is treated with the $(NH_4)_2S_x$ passivant.
○ etched in $H_2SO_4 : H_2O_2 : H_2O = 3:1:1$, ● $(NH_4)_2S_x$ passivated.

## 4.1 CONCLUSION

Experimental observations have demonstrated that simple procedures such as nitrogenation and ammonium sulphide passivation can be used to reduce the reverse leakage current of a (100) LPE GaAs Schottky diode. However, additional experimentation is essential into long term stability, the strength of the arsenic-sulphur covalent bond by exposure to device to intense light and the impermeability of the passivant to water vapour.